# Direct TEM observation and quantification of the Gibbs-Thomson effect in a nickel superalloy

C. Papadaki[a], Wei Li[b], C.S. Allen[c], M. Danaie[c], L. R. Brandt[a], A.M. Korsunsky[a]

[a] *Department of Engineering Science, University of Oxford, Parks Road, Oxford OX1 3PJ, UK*
[b] *Rolls-Royce plc, PO Box 31, Derby DE24 8BJ, UK*
[c] *Department of Materials, University of Oxford, Oxford, UK*
*Electron Physical Sciences Imaging Centre (ePSIC), Diamond Light Source, Oxford, UK*

## Abstract

Gibbs-Thompson effect is the general term referring to the influence of interfaces on the course of phase transformations such as precipitation or solidification. Whilst attention is most often focused on the Gibbs-Thomson effect on nucleation, growth and coarsening, the present study considers the reverse process of precipitate dissolution in a nickel-base superalloy during *in situ* TEM observation. The presence of several distinct populations of gamma-prime precipitates – primary, secondary, tertiary and grain boundary – allows the differences due to particle size to be quantified and interpreted. Important implications arise for the selection of heat treatment schedules for nickel-base superalloys and other alloy systems.

## Introduction

During operation, gas turbine components are subjected to extreme conditions combining significant mechanical stresses with high temperatures. To improve the overall efficiency, increasing turbine entry temperature is required that represents a challenge of expanding the operational limits of superalloys outside the current envelope. The mechanical and thermal stability at elevated temperatures and oxidative environments is one among the most crucial operational requirements for nickel-base superalloys. However, at these elevated temperatures diffusion-controlled phenomena are inevitably activated, mediating several degradation mechanisms to occur. Among them, the diffusion-controlled coarsening of the γ' precipitates can lead to the dissolution of fine particles strongly affecting the creep resistance [1]. Degradation due to diffusion-controlled phenomena may be detrimental to the lifetime of the components. To quantify and control the diffusion-induced damage, there is a dire need for detailed information about the microstructural modifications at elevated temperatures with particular attention on the dissolution of γ' precipitate populations. In this study, the microstructural evolution during heating of a new polycrystalline nickel-base superalloy is explored by performing *in situ* heating of thin lamella samples under the electron beam in a scanning transmission electron microscope (STEM). Along with the microstructure, the elemental distribution and the compositional profile of γ' precipitate populations is investigated using a synchronous electron energy loss spectroscopy (EELS) analysis. *In situ* heating experiments within STEM combined with EELS technique enables probing the mechanisms underlying the complex physicochemical processes at the nanometric scale as they unfold. *Operando* nanoscale observation of the dissolution of γ' populations was achieved, and the findings provide a direct observation of the Gibbs-Thomson effect in the dissolution behaviour of different γ' populations.

    The quantitative data included in this work describe the diffusional phenomena that take place when heating TEM lamella specimens with local thickness less than 100 nm and can provide quantitative input data for validating accurate numerical simulations. It should be noted that the diffusional phenomena observed may appear accelerated due to the small interaction volume and reduced diffusion paths within thin specimens [2]. The dwell time at each temperature level was chosen accordingly.

## Methods

The material investigated in this study is a new high γ' content polycrystalline nickel-base alloy, referred to as Alloy 11 below, which belongs to a family of new alloys that has been introduced recently. Its composition is provided in **Table 1**. A detailed description of the new family of alloys can be found in European Patent Specification EP2894234B1[3]. The material analysed was obtained from a pancake forging of Alloy 11, produced by isothermally forging a powder isostatic pressed cylindrical compact.

**Table 1:** Chemical composition of Alloy 11 (wt %)[3].

| Ni | Co | Cr | Ta | W | Al | Ti | Mo | Nb | Fe | Mn | Si | Zr | C | Bo |
|---|---|---|---|---|---|---|---|---|---|---|---|---|---|---|
| bal. | 15.06 | 12.69 | 4.77 | 3.22 | 3.16 | 2.84 | 2.14 | 1.44 | 0.95 | 0.48 | 0.47 | 0.057 | 0.027 | 0.023 |

Thin TEM lamellae samples of the material in the as-forged condition prior to any heat treatment were prepared with the site-specific lift-out technique using a LYRA3 Focused Ion Beam-Scanning Electron Microscope (FIB-SEM) system (Tescan, Brno, Czech Republic) equipped with a nanomanipulator and a gas injection system housed in the Multi Beam Laboratory for Engineering Microscopy (MBLEM, Oxford). The lifted lamella was positioned on a DENSsolutions heating chip mounted on a 45° pre-tilted support, with careful adjustment of the angle and position of the sample stage. The heating chip consists of a metal heater embedded in a chemically inert and mechanically robust silicon nitride supporting film of thickness roughly equal to 20 nm [4], on which precisely designed holes allow TEM imaging of the sample mounted on top of them. The control method used is four-point resistive feedback providing temperature accuracy < 5% [4]. The initial calibration was done by the manufacturer and calibration values are provided to be introduced in the control software, which is also used to record the temperature during the experiment. Uniform temperature distribution in the region shown in Figure 1 (b) has been verified previously with numerical simulations in [5].

To achieve the required thickness while minimising the FIB induced damage, an optimised thinning technique was employed. with lower beam-currents during the final steps. Low-energy milling while adjusting the tilting angles vastly improved the quality of the lamellae. Detailed description of the step-by-step thinning methodology for the sample preparation is presented in [11].

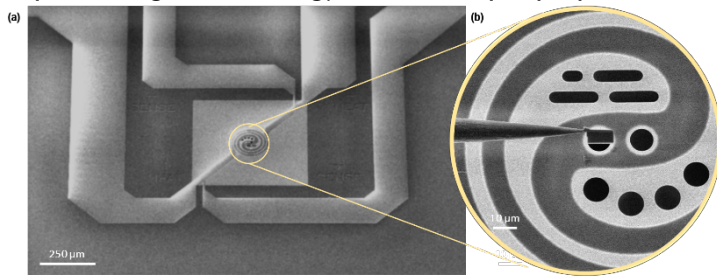

**Figure 1: (a)** SEM image of the DENSsolutions Nano-Chip used for is situ heating, **(b)** Higher magnification image of the metal spiral on the micro-hotplate of the heating chip with windows for FIB samples.

STEM studies were carried out using the probe corrected JEOL ARM200CF scanning transmission electron microscope, operated at 200 kV at the electron Physical Science Imaging Centre (ePSIC) at Diamond Light Source, Harwell Campus. For imaging of the variations of the microstructure with temperature, the bright field (BF), annular dark field (ADF) and high-angle annular dark field (HAADF) detectors were utilised. A condenser lens aperture size of 30 μm was used, while the camera length was set to 3cm. *In situ* heating of the sample was performed using the single tilt DENSsolutions heating holder. The analysis was coupled with simultaneous EELS analysis for synchronous compositional information and elemental mapping at the nanometre scale at elevated temperatures. STEM-EELS was carried out using a Gatan Quantum Dual EELS spectrometer, with collection semi-angle equal to 55.56 mrad. For the data analysis of the EELS spectrums, the Gatan Microscopy Suite (GMS) software (version 3.4.2) was utilised for the qualitative and quantitative examination of the microstructural features.

For the data analysis and quantification of the EELS measurements, a model-based quantification approach was employed using the GMS (3.4.2) in which model-based quantification has been simplified and vastly improved (e.g., [12]). including improved tolerance to the sample thickness compared to conventional quantification [13]. Plural scattering effects were incorporated into the analysis to improve accuracy, while the analysis of the electron loss near-edge fine-structure (ELNES) features was excluded.

## Experimental Results

The microstructure at room temperature of the alloy under examination is depicted in the back-scattered electron (BSE) micrograph in **Figure 2** (a). A trimodal γ′ precipitate distribution can be observed with large primary γ′ precipitates residing at the grain boundary regions, and fine secondary and tertiary γ′ precipitates distributed in the grain interior. An estimation of the size distribution of each of the populations of the γ′ precipitates was obtained with subsequent image analysis on the BSE micrographs using ImageJ software (version 1.50e). The average size of each γ′ precipitate population was estimated from the square root of the measured precipitate area. The precipitate populations appeared to conform to a trimodal size distribution, validating the presence of primary, secondary and tertiary γ′ particles. **Figure 2** (b) presents histograms of the probability distribution of the size of γ′ precipitate populations. The continuous lines denote a fit to the data: for primary γ′ a Gaussian fit was used, and the size distribution of secondary and tertiary γ′ precipitate populations was fitted with a double-Gaussian function, where the left peak corresponds to the size distribution of tertiary γ′, and the right peak to that of secondary γ′ precipitates. The table in **Figure 2** (c) summarises the mean diameters and standard deviations for each population.

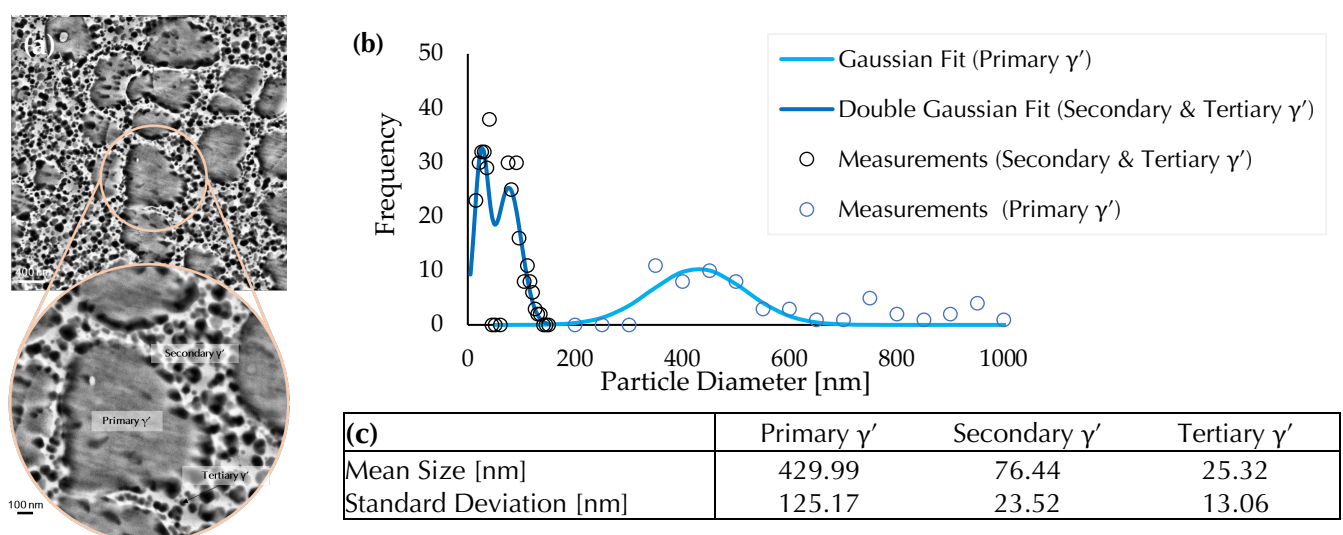

| (c) | Primary γ′ | Secondary γ′ | Tertiary γ′ |
|---|---|---|---|
| Mean Size [nm] | 429.99 | 76.44 | 25.32 |
| Standard Deviation [nm] | 125.17 | 23.52 | 13.06 |

**Figure 2: (a)** BSE images of the microstructure at room temperature. **(b)** Histograms of γ′ precipitate size distribution corresponding to secondary and tertiary γ′ precipitate populations and **(c)** Mean size and standard deviation of γ′ precipitate populations.

In situ heating of thin lamella samples in the as forged condition was performed, aiming to track the microstructural changes until the dissolution of the γ′ precipitate populations. The thin TEM samples were in situ annealed for approximately 20 minutes at different temperatures in steps of 50 °C from room temperature to a maximum temperature of 1150 °C. For every temperature increment, elemental information of the region of interest was collected by performing simultaneous EELS analysis. At elevated temperatures, EELS compositional maps can provide a sharper depiction of the microstructural features compared to the ADF images, with the latter being strongly affected by the inevitable oxidation of the sample, hindering the observation of fine γ′ precipitate populations.

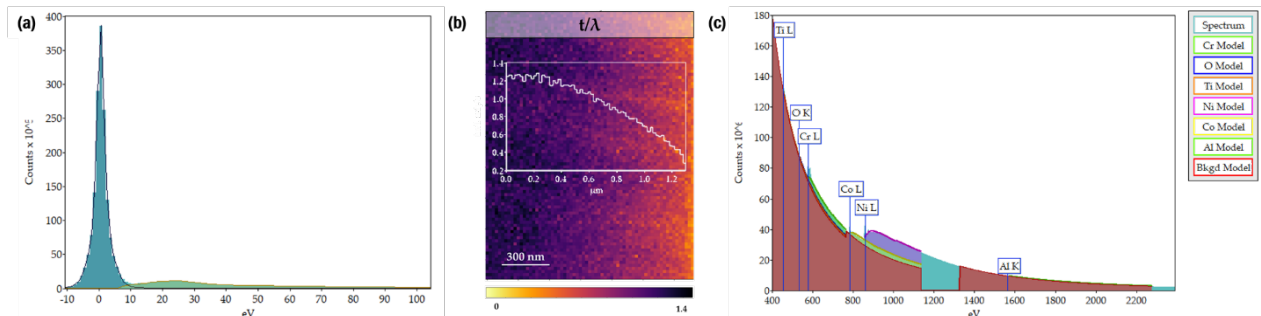

**Figure 3: (a)** Low-loss EELS spectrum at room temperature, **(b)** Map of relative thickness t/λ, **(c)** High-loss EELS spectrum at room temperature.

Apart from the qualitative examination of the microstructural features, the quantification of their elemental concentrations was performed to capture accurately the distribution of solutes and the compositional profile of the γ' precipitate populations at high temperatures. In Figure 3(a), the low-loss EELS spectrum at room temperature is shown, including the zero-loss and plasmon peaks. In the specimen under examination, the relative thickness t/λ is less than 1.2 times the inelastic mean free path, therefore the thickness is considered acceptable for quantification. However, the thickness of the sample is not uniform, as it includes a decreasing gradient towards the edge (Figure 3(b)). To avoid any errors and artifacts introduced due to the varying thickness the analysis is focused on the central region with uniform relative thickness (t/λ ≈ 1). Figure 3(b) displays the high-loss EELS spectrum at room temperature, in which the ionisation edges of the elements detected are labelled. The edges of Ti, Cr, Co, Ni and Al were selected for elemental quantification for temperatures up to 950°C. For edges Cr -L, Co-L and Ni-L, a strong peak was detected compared to the weak Al-K edge, for which the poor signal-to-noise ratio may introduce errors, and thus should be interpreted with caution. To obtain quantitative information, the cross-sections were computed using the Hartree-Slater model, which yields a good fit to the experimental data. It should be highlighted though that the main objective of the quantification in this work is to detect relative differences in the elemental concentrations while heating. However, although not rigorous, the measurements in terms of atomic percent (at. %) provide valuable estimations for the evolution of the elemental distributions while heating.

The compositional information obtained by EELS during *in situ* heating is illustrated in the elemental maps shown in Figure 4. The arrow at the top of the figure denotes the temperature which each elemental map refers to. The region of interest indicated with a dashed square in the elemental maps at room temperature contains representative precipitates of all the different γ' precipitate populations, namely primary, secondary and tertiary γ', and includes part of a grain boundary located at the bottom right corner. The range of the atomic concentration (at. %) for each element is shown in the colour bar displayed next to the corresponding elemental maps. The analysis of the same location enables direct observation of the compositional modifications during in situ heating and variations between the populations with temperature and exposure time.

Regarding tertiary γ' precipitates, the particles of this population are enveloped on all sides by the surrounding γ matrix, since the thickness of the sample is larger than the mean diameter of tertiary γ' (≈ 25 nm). This complicates the distinction between the composition of the precipitates from that of the surrounding matrix. For the accurate quantification of tertiary γ', an electrochemical method for TEM sample preparation should be employed in order to extract individual γ' precipitates by dissolving the surrounding γ matrix, as demonstrated in previous studies, such as in [8]. However, these approaches lie outside the scope of the current work which is deliberately focused on the compositional evolution while heating rather that the compositional contrast between single particles.

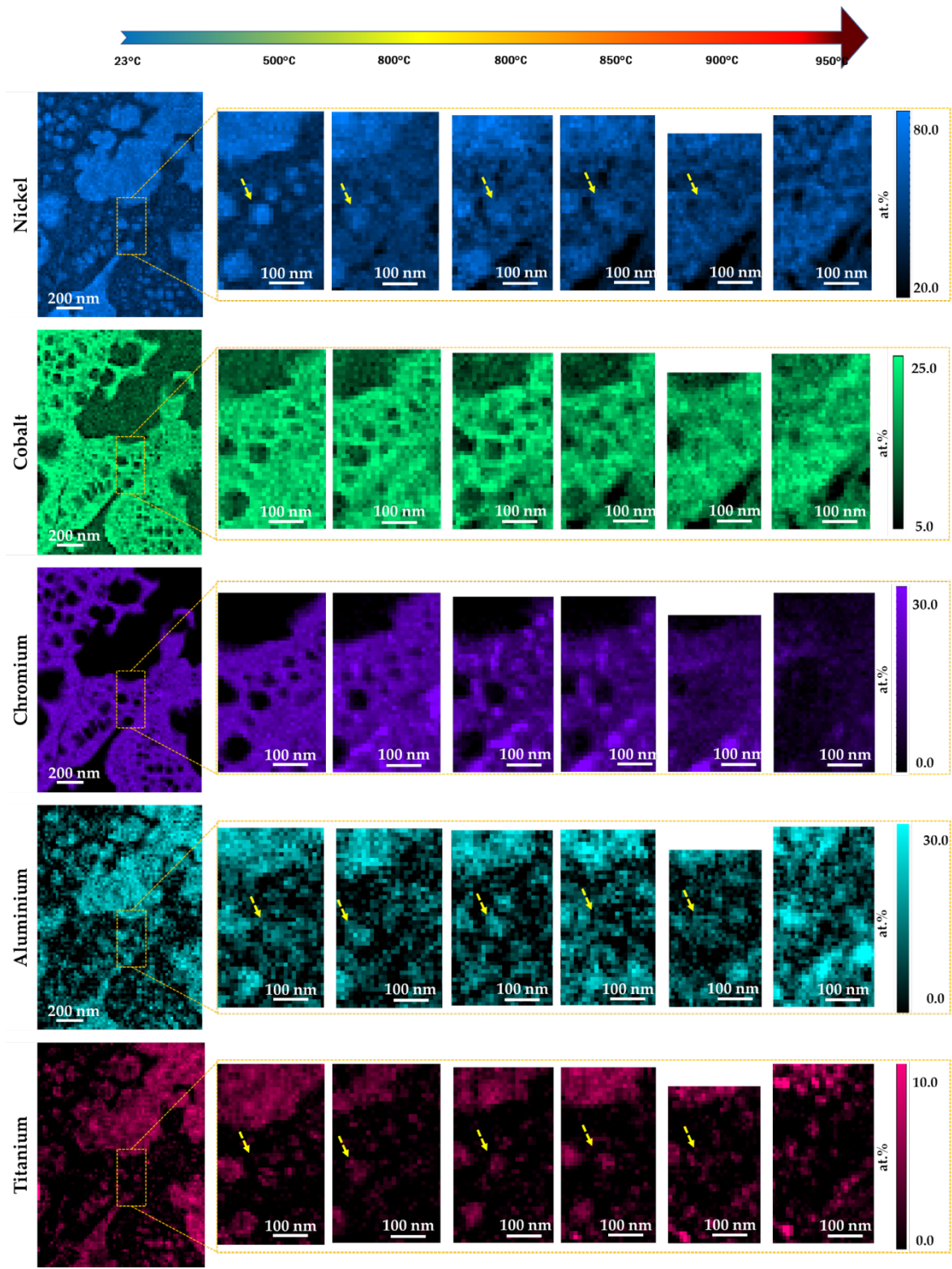

**Figure 4:** EELS elemental maps while *in situ* STEM heating, corresponding to the Ni L, Co L, Cr L, Al K and Ti L ionisation edges in the EELS spectrum.

At temperatures below 500 °C and even up to 700 °C (not included in Figure 4), no apparent changes in the elemental distributions seem to occur. At 800 °C, two successive maps were obtained, with a dwell of 20 minutes. During thermal exposure at 800 °C there is evident diffusion of chromium solutes towards the grain boundary seen in the depiction of grain boundaries in the Ni and Co maps. It is interesting to note that at this temperature the crowd of fine tertiary γ′ precipitates at the top corner of the region of interest begins to fade, especially in terms of its Ni content, indicating the initiation of the dissolution of this population. This is not observed for primary and secondary γ′ precipitates. At 850 °C, the enrichment of chromium in the grain boundary region continues, as well as Ni and Co depletion in the same region. The diffusional phenomena that occur at this high temperature trigger the gradual dissolution of secondary γ' precipitates that escalates at 900 °C. As shown by the arrows in the elemental maps of nickel and aluminium, the secondary γ' precipitates located in the proximity of the grain boundaries have vanished at 900 °C, while the tertiary γ' precipitate population appears to be completely dissolved. At even higher temperatures, at 950 °C, the large primary γ' particle at the top right corner remains undissolved, while the fine secondary and tertiary γ' populations have disappeared.

## Dissolution Sequence of γ' Precipitate Populations

The dissolution of tertiary γ' particles was experimentally observed to initiate at lower temperatures than secondary γ' particles. But how can one justify and explain this outstanding observation?

Regarding their average radii, the radius of tertiary γ' particles, $R_{0_t}$, is smaller than that of secondary, $R_{0_s}$, so $R_{0_t} < R_{0_s}$. Depending on their radii, the Gibbs–Thomson effect alters the interfacial concentration at a particle boundary. For a spherical particle precipitated in a dilute binary solid solution matrix, according to the well-known Gibbs–Thomson effect, the interfacial concentration $C_I$ is related to the curvature of particle, as follows:

$$C_I = C_e \exp\left[\frac{2\gamma\Omega}{R_B T}\frac{1}{r}\right] \tag{1}$$

where $C_e$ is equilibrium interfacial concentration, $\gamma$ is the specific interfacial energy of the matrix with the precipitate, $\Omega$ is the mean atomic (or molar) volume of the particle, $R_B$ is the Universal gas constant [8.314x103 J /(K. kmol)], $T$ is the absolute temperature and $r$ is the current radius of a spherical particle. The Gibbs–Thomson relation in Eq. (1) could be written as:

$$C_I = C_e \exp\left[\frac{2}{3}\frac{\gamma\Omega}{R_B T}S\right] \tag{2}$$

where $S$ is the surface-to-volume ratio. Since the surface-to-volume ratio of tertiary γ' precipitates is larger compared to secondary γ', the concentration at the interface of the former, $C_{I_t}$, is higher than that of the latter, $C_{I_s}$.

The dissolution of second phase particles in precipitation-hardened alloy systems is a diffusion-controlled transformation process dominated by the diffusion of solute atoms across the interface between the particles and the matrix [7]. For an isolated particle in an infinite matrix, the concentration field $C(r,t)$ of an element in solid solution can be described using Fick's second law of diffusion given by:

$$D \nabla^2 C = \frac{\partial C}{\partial t} \tag{3}$$

where $D$ is the volume diffusion coefficient of solute in the matrix. Eq. (3) must conform to the boundary conditions [9][10], with respect to the solute concentrations in the matrix, $C_M$, and at the precipitate/matrix interface, $C_I$, expressed as:

$$\begin{aligned} C(r = \infty, t) &= C_M & t \geq 0 \\ C(r, t = 0) &= C_M & r \geq R_0 \\ C(r = R_0, t) &= C_I & t > 0 \end{aligned} \tag{4}$$

In the case of dissolution, one more initial condition should be considered, which is the initial, finite, nonzero precipitate radius $R_0$:

$$r(t = 0) = R_0, \; R_0 > 0 \tag{5}$$

What is more, the flux balance condition should be satisfied as follows:

$$(C_P - C_I)\frac{dr}{dt} = D\frac{\partial C}{\partial r}\bigg|_{r=R_0} \tag{6}$$

where $C_P$, the solute concentration in the precipitate, is taken to be constant and independent of $r$ and $t$.

Exact analytical solutions for the kinetics of the diffusion-controlled phase transformations are often difficult to obtain, but several mathematical approximations have been developed. Aaron and Kotler [9] performed a detailed theoretical study in which they reviewed several widely used mathematical approximations for modelling the dissolution of spherical and planar precipitates. Namely, in their study, they compared and evaluated the reversed-growth, the invariant-field ($\partial C/\partial t = 0$) and the invariant-size ($dr/dt = 0$) analyses. In accordance with the reversed-growth approach, the time evolution of the radius of spheroidal particles $r(t)$ under isothermal conditions is given by:

$$r(t) = [R_0^2 - kDt]^{-1/2} \tag{7}$$

where $R_0$ is the initial, nonzero radius, $D$ is the volume diffusion coefficient in the matrix and $k$ is the supersaturation parameter. The supersaturation parameter, $k$, is a critical parameter describing the kinetics of the diffusional phase transformation of dissolution, given by:

$$k = 2\frac{C_I - C_M}{C_P - C_I} \tag{8}$$

It is worth pointing out that for most alloy systems $|k| < 0.3$ and in fact $|k| < 0.1$ is quite typical. In the limit of small $k$, the reversed-growth solution can provide an accurate approximation of the dissolution of spherical particles [9][10] and will be used in the context of this discussion.

For the kinetics of dissolution, $k$ is a critical parameter strongly related to the rate of dissolution in the reverse growth approximation given by Eq. (6). Assuming that both secondary and tertiary populations have the same concentration $C_P$, and since $C_{I_t} > C_{I_s}$ due to the Gibbs–Thomson effect, the value of the supersaturation parameter of tertiary γ′, $k_t$, given by Eq. (8), exceeds that of secondary γ′ precipitates $k_s$ ($C_M$ is the same for both):

$$C_{I_t} > C_{I_s} \Rightarrow k_t > k_s \tag{9}$$

Therefore, since $k_t > k_s$ and $R_{0_t} < R_{0_s}$, it can be concluded that under isothermal conditions less time is required for the dissolution of smaller tertiary γ′ precipitates than for larger secondary γ′ particles, according to the reverse growth approximation (Eq. (7)).

Considering a binary system under increasing temperature, the dissolution of second phase particles is energetically driven in accordance with the Gibbs energy minimisation principle. The total free energy change includes the contributions of the volume free energy $\Delta G_V$, the interfacial energy per unit area, $\gamma$, and the lattice distortion of the matrix. Progression towards equilibrium is accomplished by a decrease in the interfacial energy term achieved by particle dissolution when the temperature exceeds the solvus temperature $T_s$. When the temperature is equal to $T_s$, both the matrix and the secondary phases are thermodynamically stable and they have the same free energy, so:

$$\Delta G = 0 \Leftrightarrow \Delta H - T_s \Delta S = 0 \Leftrightarrow \Delta H = T_s \Delta S \tag{10}$$

Since the difference between specific phase entropies vary slowly around the solvus temperature, it is often assumed to be constant and equal to its value $\Delta S$ at $T_s$. Then the temperature dependence of $\Delta G$ can be expressed as:

$$\Delta G(T) = \Delta H - T \Delta S = T_s \Delta S - T \Delta S \Rightarrow \Delta G(T) = (T_s - T) \Delta S \tag{11}$$

Defining $\Delta T = T - T_s$ as the degree of overheating, one concludes that as the temperature increases over the value of the solvus temperature $T_s$, the thermodynamic driving force for the dissolution of the secondary phase increases. Additionally, the diffusion rate increases with increasing temperature following an Arrhenius type of temperature dependence that leads to variation over several orders of magnitude, but over large range of temperatures. Considering the above, in the vicinity of the solvus temperature, $T_s$, the temperature dependence of the rate of dissolution could be assumed to be linearly related to the degree of overheating. This could be expressed by the following form in terms of the supersaturation parameter, $k$:

$$k \propto (T - T_s) \tag{12}$$

since the rate of dissolution is correlated to $k$ for the kinetics of dissolution. Therefore, due to the different supersaturation parameters, the dissolution of tertiary γ′ should occur at lower temperatures than for secondary γ′ particles during heating:

$$k_t > k_s \Rightarrow T - T_{s_t} > T - T_{s_s} \Rightarrow T_{s_t} < T_{s_s} \tag{13}$$

This conclusion is in accordance with the experimental observations described in this study. The dissolution of finer particles (tertiary γ′) was observed to initiate at lower temperatures than larger sized particles (secondary γ′). This observation has been explained using basic thermodynamic principles and diffusion kinetics. Conforming to the Gibbs-Thomson relation, the difference between $C_{I_t}$ and $C_{I_s}$ induces a diffusive flux of solute atoms from smaller to larger particles. Solutes diffuse through the concentration gradients across the surface of the smaller sized particles through the matrix to the surface of larger sized particles. Likewise, in diffusion-driven coarsening the average particle radius increases, and the total number of particles decreases with time, as does the total free surface energy of the system resulting in the reduction of Gibbs free energy [6].

## Discussion & Conclusions

STEM mapping yields spatially resolved information from nano- down to atomic-scale, and with the help of elemental concentration quantification using EELS can reveal the evolution of diffusional phenomena during heating. It has been shown that model-based quantification of EELS measurements offers a reasonably robust way of performing compositional quantification [12] [13]. The principal challenge concerns achieving good accuracy, since artefacts and errors may be introduced, e.g., due to incorrect determination of background for subtraction, of signal sum width, etc. Several modifications and extensions of this approach have been suggested to obtain a more accurate quantification [13]. However, this would solely affect the absolute values of the compositional measurements and not alter the elemental distributions and the overall trends in terms of evolution with temperature for the different elements. Therefore, the results obtained are undoubtedly reliable in terms of compositional variations and elemental distributions, while the absolute values should be treated as rough estimations of the actual atomic composition and not as precise measurements.

*In situ* heating under STEM observation provides a valuable method for the observation of each population separately and enables direct comparisons between them. Other experimental approaches such as thermo-analysis techniques and neutron diffraction can be employed to study the evolution of the γ′ precipitate populations in terms of structure and misfit strain with the γ matrix while heating in γ-γ′ alloy systems []. However, the similarity in lattice parameters between the secondary and tertiary γ′ populations leads to tightly overlapping diffraction peaks which prevents reliable fitting of them separately. Therefore, although STEM-EELS provides poor statistics and may not be representative of the bulk behaviour, it displays a competitive advantage offering unparalleled local information. Using this technique, it has been possible to obtain experimental evidence of the difference in the solvus temperature between different γ′ precipitate populations. The local STEM-EELS data provides first direct evidence of the different behaviour of secondary and tertiary γ′ populations with temperature in terms of dissolution.

This variation has also been observed and reported previously. Masoumi et al. [6] studied the mechanisms and kinetics of the reprecipitation of γ′ particles during cooling from super- and sub-solvus temperatures in alloy AD730™, a nickel-base superalloy for turbine disc applications. Using Differential Thermal Analysis (DTA) they detected two distinct endothermal peaks while heating corresponding to the dissolution of secondary and primary γ′ precipitate populations. In particular, depending on the heating rate, the dissolution of secondary γ′ particles occurred in the range between 800 to 825 °C for heating rates equal to 65 °C/min and 120 °C/min respectively. The dissolution of primary γ′ happened around 1080 °C for the 10 °C/min heating rate and 1120 °C for the 120 °C/min heating rate. These observations are in good agreement with our approach, considering the larger size of primary compared to secondary γ′ particles.

# Acknowledgements

This research was funded by Rolls-Royce plc through a grant to the Oxford University Technology Centre (UTC) in Solid Mechanics. The material analysed was supplied by Rolls-Royce plc.